\documentclass[conference]{IEEEtran}
\IEEEoverridecommandlockouts
\usepackage{cite}
\usepackage{amsmath,amssymb,amsfonts}
\usepackage{algorithmic}
\usepackage{graphicx}
\usepackage{textcomp}
\usepackage{xcolor}
\usepackage{multirow}
\usepackage{comment}
\usepackage{booktabs}

\def\BibTeX{{\rm B\kern-.05em{\sc i\kern-.025em b}\kern-.08em
    T\kern-.1667em\lower.7ex\hbox{E}\kern-.125emX}}
\begin{document}

\title{Speech Enhancement Using Continuous Embeddings of Neural Audio Codec \\

\thanks{
\textsuperscript{*}equal contribution}
}

\makeatletter
\newcommand{\linebreakand}{%
  \end{@IEEEauthorhalign}
  \hfill\mbox{}\par
  \mbox{}\hfill\begin{@IEEEauthorhalign}
}
\makeatother

\author{\IEEEauthorblockN{Haoyang Li\textnormal{*}}
\IEEEauthorblockA{\textit{Nanyang Technological University}\\
Singapore \\
li0078ng@e.ntu.edu.sg}
\and
\IEEEauthorblockN{Jia Qi Yip\textnormal{*}}
\IEEEauthorblockA{\textit{Nanyang Technological University}\\
Singapore \\
jiaqi006@e.ntu.edu.sg}
\linebreakand 
\IEEEauthorblockN{Tianyu Fan}
\IEEEauthorblockA{\textit{Nanyang Technological University}\\
Singapore\\
fant0004@e.ntu.edu.sg}
\and
\IEEEauthorblockN{Eng Siong Chng}
\IEEEauthorblockA{
\textit{Nanyang Technological University}\\
Singapore \\
aseschng@ntu.edu.sg}
}

\maketitle

\begin{abstract}
Recent advancements in Neural Audio Codec (NAC) models have inspired their use in various speech processing tasks, including speech enhancement (SE). In this work, we propose a novel, efficient SE approach by leveraging the pre-quantization output of a pretrained NAC encoder. Unlike prior NAC-based SE methods, which process discrete speech tokens using Language Models (LMs), we perform SE within the continuous embedding space of the pretrained NAC, which is highly compressed along the time dimension for efficient representation. Our lightweight SE model, optimized through an embedding-level loss, delivers results comparable to SE baselines trained on larger datasets, with a significantly lower real-time factor of 0.005. Additionally, our method achieves a low GMAC of 3.94, reducing complexity 18-fold compared to Sepformer in a simulated cloud-based audio transmission environment. This work highlights a new, efficient NAC-based SE solution, particularly suitable for cloud applications where NAC is used to compress audio before transmission.
\end{abstract}


\section{Introduction}
\label{sec:intro}
Speech enhancement (SE) is the process of improving the quality and intelligibility of speech signals by reducing noise and distortion, benefiting applications in telecommunications, hearing aids and speech recognition. Deep learning-based speech enhancement (SE) methods, which aim to estimate clean speech from noisy speech, utilize architectures such as convolutional neural networks \cite{fu2018end} \cite{chen2021time}, recurrent neural networks \cite{zhao2018convolutional} and Transformers \cite{yu2022dual} \cite{yu2022setransformer} to learn complex patterns and adapt to various noise environments, outperforming traditional techniques. 

Deep learning SE methods can be broadly separated into discriminative methods and generative methods. Discriminative methods aim to learn a mapping between noisy and clean speech through methods such as time frequency masking \cite{chakrabarty2018time} and complex spectral mapping \cite{fu2017complex}. However, these supervised approaches cannot ensure generalization to all acoustic scenarios due to the restricted diversity of the training data, and can even introduce undesirable speech distortions \cite{wang2019bridging}. On the other hand, generative methods capture the underlying distribution of clean speech, using it as a prior to guide the enhancement process. These methods utilize techniques such as flow \cite{nugraha2020flow}, variational autoencoders \cite{fang2021variational} and diffusion \cite{hu2023noise} \cite{lu2022conditional}, and have shown to generalize better to out-of-domain noise than discriminative methods \cite{fang2021variational} \cite{lu2022conditional}.

The goal of a Neural Audio Codec (NAC) is to compress a segment of audio using codebooks to allow for memory-efficient transmission of the audio signal, typically between a local and remote or cloud device. Recent advancements in NAC models \cite{zeghidour2021soundstream} \cite{defossez2022high} \cite{kumar2024high} have used large amounts of data to pre-train models that can retain high fidelity despite very significant compression. This has inspired the use of these NACs for downstream tasks such as text-to-speech \cite{wang2023neural}, speech separation~\cite{yip2024towards} and generative SE \cite{wang2024speechx} \cite{yang2023uniaudio}. Current NAC-based SE methods utilize the speech tokens from the NAC models' codebooks to formulate SE as token prediction tasks where the objective is to predict the NAC tokens of target clean speech through Language Models (LM). While LM-based SE methods have shown promising enhancement results, they can be computationally expensive depending on the LM used. For example, Uniaudio \cite{yang2023uniaudio} predicts clean speech tokens through autoregressive LM and is known to have a slow Real-Time Factor (RTF) of around 10 on an A100 GPU \cite{li2024masksr}. In contrast with autoregressive LM-based SE methods \cite{wang2024speechx} \cite{yang2023uniaudio}, MaskGit style LM-based SE \cite{li2024masksr} \cite{yang2024genhancer} are recently introduced which speeds up inference by predicting multiple tokens concurrently.

In this work, we propose a novel NAC-based solution for SE without utilizing LM, by performing SE directly using the pre-quantization output of a NAC encoder with a light-weighted SE model. Secondly, we propose an embedding-level loss function to train this NAC-based SE model. Due to the high compression ratio in the time dimension, the loss function requires less computation, allowing for faster training. Our method is also highly efficient over non-NAC based SE models in cloud applications where audio codec is typically used to compress audio signal before transmission \cite{defossez2022high} \cite{rao1996techniques} (see figure \ref{fig:overview} for illustration). Our method achieves a low Real Time factor of 0.005 while maintaining comparable speech quality to previous SE works trained on identical or larger datasets. To our knowledge, we are the first to perform SE on the intermediate features of a NAC without utilizing LM.

\section{Methodology}

\subsection{Overall Architecture}
As illustrated in figure \ref{fig:overall_arch}a, our SE architecture consists of a pretrained codec encoder, a light-weighted SE module, and a pretrained codec quantizer and decoder. Noisy speech $y_{in}$ first undergoes encoding through the DAC encoder to obtain a continuous representation $y_{e}$. SE is then performed directly on the encoded representation $y_{e}$ to obtain $y_{h}$, the predicted continuous embedding of the target clean speech. The enhanced representation is then fed into the codec's quantizer module to obtain the quantized representation $y_{h}^{q}$, which can be fed into the codec decoder to generate the enhanced speech $y_{out}$. More concretely,
\begin{equation}
    \begin{aligned}
    y_{e} &= \text{DACEncoder}(y_{in})\\
    y_{h} &= \text{SEModel}(y_{e})\\
    y_{h}^{q} &= \text{Quantizer}(y_{h}) \\
    y_{out} &= \text{DACDecoder}(y_{h}^{q})  \\
    \end{aligned}
\end{equation} 

\noindent for $y_{in}, y_{out}\in\mathbb R^{L}$ and $y_{e}, y_{h}, y_{h}^{q}\in{}\mathbb R^{D\times T}$ where $L$ is the waveform length, $D$ is size of the DAC embedding dimension and $T$ is the time dimension after compression by DAC.

\subsection{Neural Audio Codec}
The neural audio codec used in this work is the Descript Audio Codec (DAC)~\cite{kumar2024high}, a high fidelity universal codec for speech, music and environmental sounds. To match the 16kHz audio used in our SE experiments, we use the 16kHz version of the codec with pretrained weights provided by the original authors\footnote{https://huggingface.co/descript}. During training, the codec is kept frozen. The codec's encoder compresses input audio by 320 times along the time dimension (i.e. $T=\frac{L}{320}$) while having a channel size of 1024 (i.e. $D=1024$). The encoding significantly improves the computational efficiency for the downstream SE task by reducing sequential operations along the time dimension while leveraging parallel processing across the channel dimension. 

\subsection{SE Module}
As illustrated in figure \ref{fig:overall_arch}b, the SE module is located between the DAC encoder and the DAC quantizer, and is trained to perform enhancement on the encoded noisy speech $y_{e}$. Similar to~\cite{yip2024towards}, the bulk of the SE module consists of a series of transformer layers. The Modulation Block consists of two parallel Conv1D pathways: one applies a sigmoid activation to act as a gating mechanism, while the other uses a Snake1D activation \cite{ziyin2020neural} to introduce non-linear feature transformations. The output of both branches are multiplied and passed through another Snake1D activation, enabling dynamic modulation of feature representations for enhanced refinement. For speech enhancement, we modified the output layers to produce only a single set of enhanced output embeddings. 

\begin{figure}[!t]
    \includegraphics[scale=0.65]{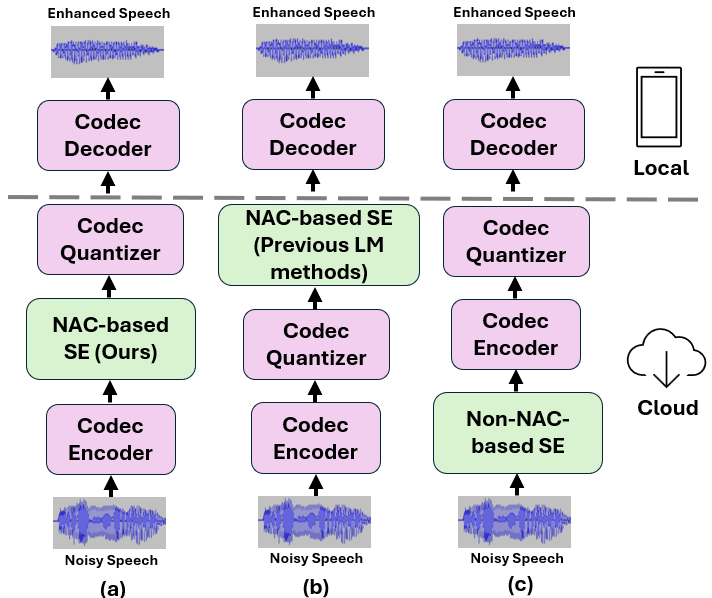}
    \caption{Overview of different speech enhancement pipelines depicting scenarios where speech is processed on the cloud and transmitted to a local device. 
    (a) Our approach (b) Previous NAC-based SE solutions leveraging LM \cite{wang2024speechx} \cite{yang2024genhancer} (c) SE methods that do not utilize NAC}
    \label{fig:overview} 
\end{figure}

\subsection{Loss Functions}
\subsubsection{Embedding Loss Function}
To guide the learning of the SE model in the latent space of the codec, we introduce an embedding loss (\ref{eq:emb_loss}) between the enhanced latent representation $y_{h}$ and the clean latent representation $x_{e}$.

\begin{equation}
\begin{aligned}
    x_{e} &= \text{DACEncoder}(x_{in}) \\
    L_{\text{Emb}} &= \mathbb{E}_{{x_{e}}, {y_{h}}} \| x_{e} - y_{h} \|_1
\end{aligned}
\label{eq:emb_loss}
\end{equation}
\noindent where $x_{in}\in\mathbb R^{L}$ is the clean speech and $x_{e}\in{}\mathbb R^{D\times T}$.

\subsubsection{Time and Frequency domain Loss Functions}
To further improve the perceptual quality of the enhanced speech, we utilize a time-domain loss (\ref{eq:time_loss}) and frequency-domain loss (\ref{eq:freq_loss}) between the enhanced speech $y_{out}$ and the clean speech after transmitting through the DAC codec $x_{out}$.

\begin{equation}
\begin{aligned}
    x_{e}^{q} &= \text{Quantizer}(x_e) \\
    x_{out} &= \text{DACDecoder}(x_{e}^{q}) \\
\end{aligned}
\end{equation}
\begin{equation}
\begin{aligned}
    L_{\text{Time}} &= \mathbb{E}_{{x_{out}}, {y_{out}}}  \| x_{out} - y_{out} \|_1 
\end{aligned}
\label{eq:time_loss}
\end{equation}

\begin{equation}
\begin{aligned}
L_{\text{Freq}} &= \mathbb{E}_{{x_{out}}, {y_{out}}} \| \text{Mel}(x_{out}) - \text{Mel}(y_{out}) \|_2^2
\end{aligned}
\label{eq:freq_loss}
\end{equation}

\noindent where Mel represents the audio to MelSpectrogram transformation, $x_{out}\in\mathbb R^{L}$ and $x_{e}^{q}\in{}\mathbb R^{D\times T}$.

\subsubsection{Overall Loss Function}
The overall loss function (\ref{eq:overall_loss}) is the sum of the individual losses, where $\alpha$, $\beta$ and $\gamma$ are used to balance the impact factor of each loss, and are set to $1$, $500$ and $\frac{1}{11}$ respectively in our experiments.

\begin{equation}
\begin{aligned}
L_{\text{Overall}} = \alpha \times L_{\text{Emb}} + \beta \times L_{\text{Time}} + \gamma \times L_{\text{Freq}}
\end{aligned}
\label{eq:overall_loss}
\end{equation}

\begin{figure*}[!t]
    \centering
    \includegraphics[scale=0.55]{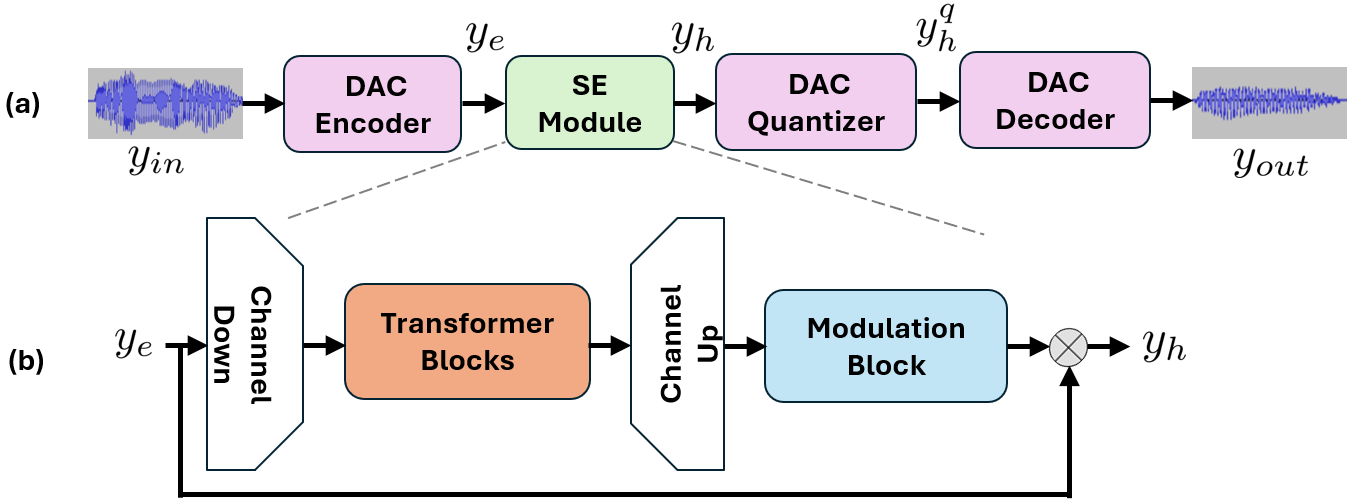}
    \caption{(a) Overall architecture of our SE pipeline 
    (b) Architecture of our embedding-domain SE model}
\label{fig:overall_arch} 
\end{figure*}

\section{Experiments}
\label{sec:Experiments}
\subsection{Dataset and Evaluation Metrics}
 $\textbf{Trainset}$ Following previous NAC \cite{li2024masksr} and LM-based \cite{wang2024selm} SE works, we gathered clean speech and noise clips to generate noisy speech on the fly during training with signal-to-noise ratio (SNR) in [-5, 20] dB. Specifically, our train set consists of 40 hours of clean speech and 15 hours of noise clips from the 2022 DNS Challenge \cite{dubey2022icassp}, where the clean speech comes from the read speech and the VCTK \cite{veaux2017cstr} partition, and the noise clips comes from the audioset and freesound partition. Our validation set is gathered in an identical manner with 4 hours of clean speech and 1.5 hours of noise.

 $\textbf{Testset}$ Following previous works \cite{li2024masksr} \cite{wang2024selm}, we use the 2020 DNS Challenge test sets \cite{reddy2020interspeech}, which include 
 2 synthetic testsets with and without reverb respectively as well as a testset of real recordings.

$\textbf{Evaluation metrics}$ 
We follow previous works \cite{li2024masksr} \cite{wang2024selm} to evaluate speech quality using the publicly available DNSMOS \cite{reddy2021dnsmos}, a robust neural network-based perceptual quality
estimator. Unlike the reference-based PESQ \cite{rix2001perceptual} metric which cannot reliably measure speech quality when misalignment between reference and target speech occurs \cite{serra2022universal} \cite{wang2024selm} \cite{li2024masksr}, DNSMOS is a learned, reference-free metric which can predict the quality ratings regardless of any potential misalignments. Additionally, we measure the speaker similarity between the enhanced speech and the target clean speech through cosine similarity calculated by x-vector embeddings \cite{snyder2018x}. Finally, we also estimate the speech intelligibility of the enhanced speech through word error rate (WER) calculated by an ASR model.

\subsection{Experimental Setup and Baselines}
All experiments are performed on a single A100 GPU with RAM size of 40GB. We trained all of our models at 16k sampling rate and learning rate of ${1.5 \times 10^{-4}}$ for 70 epochs using the Adam optimizer. Only the parameters of the SE module are updated during training. For the SE module, we set the number of the transformer blocks to 8 and the embedding size to 256. Training was done with batch size of 4.

We also trained a Sepformer model \cite{subakan2021attention} as baseline using the default recipe from Speechbrain \cite{speechbrain} on our training set. Sepformer is a transformer-based model used for speech separation \cite{subakan2021attention} and enhancement \cite{de2022efficient}. It is similar to our transformer model and serves as a suitable candidate for comparison. The main difference between our model and Sepformer is the use of DAC as the encoder and decoder, as well as the use of only a single stack of transformers, compared to Sepformer's dual-path architecture.

In Table~\ref{tab:compare_against_sep_efficiency} we compare both systems under a practical online audio transmission scenario, where the audio signal is processed on the cloud and sent to a local device (see figure \ref{fig:overview}(a) and (c) for illustration). To investigate whether the codec transmission degrades the performance of the Sepformer model, we also report the Sepformer model's performance without undergoing transmission through the DAC codec.

In addition, we use the results from \cite{wang2024selm} for CDiffuSE \cite{lu2022conditional}, SGMSE \cite{welker2022speech}, StoRM \cite{lemercier2023storm} and SELM \cite{wang2024selm} to compare with our work. These methods are generative SE methods, where the first 3 are diffusion-based methods and the last is a LM-based method. These baselines are trained by \cite{wang2024selm} using clean speech and noisy speech generated on the fly similar to our approach. It should be noted that their training dataset is gathered from more diverse sources including the DNS challenge, LibriMix, WHAM! and DEMAND, and is significantly larger than ours.

\section{Results and discussions}

\begin{table}[h!]
  \caption{Efficiency comparison between our approach and Sepformer (Cloud) on the DNS Test Set (Without Reverb).}
  \label{tab:compare_against_sep_efficiency}
  \centering
    \begin{tabular}{ccc}
    \hline
    Model & GMACs $\downarrow$ & Training time (h/epoch) $\downarrow$ \\
    \hline
    Sepformer Cloud &  69.29 & 2.70\\
    Ours & \textbf{3.94} & \textbf{0.66}\\
    \hline
    \end{tabular}
\end{table}

\begin{table}[h!]
  \caption{Speech quality comparison between our approach and Sepformer on the DNS Test Set (Without Reverb).}
  \label{tab:compare_against_sep_quality}
  \centering
    \begin{tabular}{cccccc}
    \hline
    Model & SIG $\uparrow$ & BAK $\uparrow$ & OVL $\uparrow$ & Cos. $\uparrow$ & WER $\downarrow$ \\
    \hline
    Sepformer Cloud & \textbf{3.566} & \textbf{4.134} & \textbf{3.340} & \textbf{0.989} & 0.084 \\
    Sepformer Oracle & 3.549 & 4.119 & 3.318 & 0.988 & 0.081 \\
    Ours & 3.475 & 3.880 & 3.128 & 0.985 & \textbf{0.080} \\
    \hline
    \end{tabular}
\end{table}

\begin{table*}[htbp]
    \caption{Performance comparison of our model against results reported in SELM \cite{wang2024selm}.}
    \label{tab:compare_against_sota}
    \centering
    \resizebox{0.9\textwidth}{!}{
    \resizebox{\textwidth}{!}{
    \begin{tabular}{@{}cccccccccccccc@{}}
    \toprule
    \multirow{3}{*}{System} &
      \multirow{3}{*}{RTF$\downarrow$} &
      \multicolumn{3}{c}{With Reverb} &
      \multicolumn{3}{c}{Without Reverb} &
      \multicolumn{3}{c}{Real Recordings} \\
    \cmidrule(lr){3-5} \cmidrule(lr){6-8} \cmidrule(lr){9-11}
     & & \multicolumn{3}{c}{DNSMOS $\uparrow$} & \multicolumn{3}{c}{DNSMOS $\uparrow$} &  \multicolumn{3}{c}{DNSMOS $\uparrow$} \\
    \cmidrule(lr){3-5} \cmidrule(lr){6-8} \cmidrule(lr){9-11}
     & & SIG & BAK & OVL & SIG & BAK & OVL & SIG & BAK & OVL \\
    \midrule
    \addlinespace[1ex]
    Noisy & - & 1.760 & 1.497 & 1.392 & 3.392 & 2.618 & 2.483 & 3.053 & 2.510 & 2.255 \\
    \midrule
    \addlinespace[1ex]
    CDiffuSE & 0.086 & 2.541 & 2.300 & 2.190 & 3.294 & 3.641 & 3.047 & 3.201 & 3.104 & 2.781 \\
    SGMSE & 0.675 & 2.730 & 2.741 & 2.430 & 3.501 & 3.710 & 3.137 & 3.297 & 2.894 & 2.793 \\
    StoRM & 0.994 & 2.947 & 3.141 & 2.516 & \textbf{3.514} & 3.941 & 3.205 & 3.410 & 3.379 & 2.940 \\
    SELM & - & \textbf{3.160} & \textbf{3.577} & \textbf{2.695} & 3.508 & \textbf{4.096} & \textbf{3.258} & \textbf{3.591} & 3.435 & \textbf{3.124} \\
    
    \midrule
    \addlinespace[1ex]
    
    \hspace{1em} Ours (Small-scale trainset) & \textbf{0.005} & 2.889 & 3.133 & 2.330 & 3.475 & 3.880 & 3.128 & 3.203 & \textbf{3.807} & 2.857 \\
    \bottomrule
    \end{tabular}
    }
    \label{tab:results}
    }
\end{table*}

\begin{table}[h!]
  \caption{Investigation on the impact of different loss functions}
  \label{tab:ablation}
  \centering
    \resizebox{0.45\textwidth}{!}{
    \begin{tabular}{ cccc }
    \hline
    Loss & SIG $\uparrow$ & BAK $\uparrow$ & OVL $\uparrow$ \\
    \hline
    $L_{\text{Emb}}$ & \textbf{3.544} & 3.596 & 3.052 \\
    $L_{\text{Time}} + L_{\text{Freq}}$ & 1.200 & 1.169 & 1.080
    \\
    $L_{\text{Time}} + L_{\text{Freq}} + L_{\text{Emb}}$ & 3.475 & \textbf{3.880} & \textbf{3.128} \\
    \hline
    \end{tabular}
    }
\end{table}

\subsection{Comparison with baselines}
Table~\ref{tab:compare_against_sep_efficiency} compares the efficiency of our method against Sepformer (Cloud) during training and inference. Our SE model reduces training time by ~4$\times$. This is largely because the SE module operates on the latent space of the frozen DAC model, which significantly reduces the temporal dimension of the input to the SE model in comparison with Sepformer. Similarly, we notice about 18 times reduction in MAC (Multiple and Accumulate operations) during inference. The MAC results are calculated using PyTorch-OpCounter\footnote{https://github.com/Lyken17/pytorch-OpCounter} on a 10 seconds long utterance. The MAC for audio compression and decompression are omitted since they are used by both the speech enhancement models during the online audio transmission scenario as shown in Figure \ref{fig:overview}.

In Table~\ref{tab:compare_against_sep_quality}, we compare the speech quality of our method against both Sepformer in the online transmission scenario (Sepformer Cloud) where the output of Sepformer is transmitted through the DAC codec as well as the raw output of Sepformer (Sepformer Oracle). While our method achieves performance slightly below Sepformer, it offers a substantial advantage in terms of computational efficiency, significantly reducing resource requirements while maintaining good-quality results. We also find that Sepformer (Cloud) outperforms Sepformer (Oracle) slightly in most metrics except for WER, suggesting that the transmission through the DAC codec does not degrade speech quality.

Table~\ref{tab:compare_against_sota} compares our work against several generative SE methods trained by \cite{wang2024selm}. Despite being trained on a smaller dataset, our model achieves comparable results to most of the reported SE methods. Our method, including the DAC components, achieves a Real-Time-Factor (RTF) of 0.005, which is lower than all reported baselines. For all RTF calculations, we obtained the results by performing inference on the DNS Challenge Test Set (Without Reverb) for 5 times on a single A100 GPU and averaging the results. The RTF for SELM is omitted since its implementation is not publicly available.

\subsection{Investigation on loss functions}

We also investigate the impact of different loss functions. In table \ref{tab:ablation}, we show that the SE module fails to learn without the proposed embedding loss $L_{\text{Emb}}$, indicating that $L_{\text{Emb}}$ is crucial in guiding the learning of the SE module in the embedding space of the DAC model. We also observe that adding $L_{\text{Time}}$ and $L_{\text{Freq}}$ to $L_{\text{Emb}}$ improves speech quality, indicating that direct optimization on the generated speech is useful for refining speech quality.

\section{Conclusion}
In this work, we proposed a novel and efficient approach to speech enhancement by leveraging the pre-quantization output of a neural audio codec. Our method achieves comparable speech quality to previous works while significantly reducing computational complexity, with a 18x reduction in GMACs compared to Sepformer (Cloud) and a real-time factor of 0.005. We introduced an embedding-level loss function to train the SE model and demonstrated its effectiveness through ablation studies. Our approach is particularly well-suited for cloud applications where audio compression is typically used before transmission. Future work could explore the potential of this approach with larger datasets and more complex model architectures to further improve enhancement quality while maintaining computational efficiency. We encourage readers to listen to our demo samples\footnote{https://nac-se-continuous-emb.netlify.app/}, which showcase the performance of our proposed method.

\section{Acknowledgement}
The research is supported by the National Research Foundation, Singapore (AISG Award No: AISG2-GC-2022-005), and the computational work was partially performed on resources of the National Supercomputing Centre, Singapore. Any opinion and findings in the work are solely those of the authors.

\newpage

\bibliographystyle{IEEEbib}

\bibliography{refs}

\end{document}